\documentclass[prb,twocolumn,preprintnumbers,amsmath,amssymb,showpacs,showkeys]{revtex4}

\usepackage[dvips]{color}
\usepackage{graphicx}
\usepackage{amsfonts}
\usepackage{amssymb}
\usepackage{epsfig}
\usepackage{bm}
\usepackage{dcolumn}

\newcommand{\beq}{\begin{eqnarray}}
\newcommand{\eeq}{\end{eqnarray}}
\newcommand{\e}{{\text e}}

\newcommand{\mf}{\textrm{{\scriptsize MF}}}
\newcommand{\imf}{\textrm{{\scriptsize IMF}}}
\newcommand{\sw}{\textrm{{\scriptsize SW}}}
\def\bk{k}

\def\GS{0}
\def\bphi{{\mbox{\boldmath $\phi$}}}
\def\bpsi{{\mbox{\boldmath $\psi$}}}
\def\bB{{\mbox{\boldmath $B$}}}
\def\bA{{\mbox{\boldmath $A$}}}

\begin{document}
\title{Superfluid to Mott-insulator transition of hardcore bosons in a superlattice}
\date{\today}

\author{Itay Hen}
\email{itayhe@physics.georgetown.edu}
\author{Marcos Rigol}
\email{mrigol@physics.georgetown.edu}
\affiliation{Department of Physics, Georgetown University, Washington, DC 20057, USA}

\begin{abstract}
We study the superfluid to Mott-insulator (SF-MI) transition of hardcore bosons in commensurate
superlattices in two and three dimensions. We focus on the special case where the
superlattice has period two and the system is at half filling. We obtain numerical results 
by using the stochastic series expansion (SSE) algorithm, and compute various properties 
of the system, such as the ground-state energy, the density of bosons in the zero-momentum mode, the superfluid density, 
and the compressibility. We employ finite-size scaling to extrapolate the thermodynamic limit, and find 
the critical points of the phase transition. We also explore 
the extent to which several approximate solutions such as mean-field theory, with and without 
spin-wave corrections, can help one gain analytical insight into the behavior of the 
system in the vicinity of the phase transition.
\end{abstract}


\maketitle

\section{Introduction}
Recent developments in the field of ultracold Bose gases have opened
a new promising avenue of theoretical and experimental research 
in the study of the phases of quantum matter.
A gas of bosonic atoms in an optical trap has been
reversibly tuned from a Bose-Einstein condensate to a 
state composed of localized atoms as the strength of a periodic optical potential was varied.\cite{gr1,gr2} 
This is an example of a quantum phase transition; a phase transition
generated by quantum fluctuations and correlations rather than by
a competition between the energy of a system and the entropy of its thermal fluctuations.\cite{sachdev}
Understanding this phenomenon has emerged as one of the most challenging 
and interesting tasks of condensed matter physics.
Theoretically, it is generally accepted that it 
can be studied using the Bose Hubbard model, where the transition
is thought to be from a superfluid phase to a Mott-insulator (SF-MI), as examined
in the seminal paper by Fisher {\it et al.},\cite{r4} with an application to 
${^4}$He absorbed in porous media in mind. The relevance of the Bose-Hubbard model
to gases of alkali-metal atoms in optical lattices was realized in Ref.\ \onlinecite{r5}, 
and recent developments have been reviewed in Refs.\ \onlinecite{r6} and \onlinecite{bloch08}. 

Interestingly, the Bose-Hubbard model is nonintegrable even in one dimension
(as opposed to, say, its fermionic counterpart\cite{fhb}). Gaining analytical insight into the SF-MI phase transition thus normally 
requires resorting to numerical and variational methods such as strong-coupling expansion,\cite{sce1,sce2} 
coarse graining,\cite{cg}
mean-field theories,\cite{mft} field-theoretical approaches\cite{ftapp} or other perturbative methods\cite{per1} 
for a better understanding 
of this phenomenon. Within the variational approach, the phase transition is taken to be 
the point at which the variational ansatz has lower energy than a delocalized 
Bogoliubov state (where a fixed particle number at each lattice site is constrained).

In a recent paper, Aizenman {\it et al.}\cite{Aizenman} considered an alternative model 
for the study of the SF-MI phase transition. They studied the half-filled Bose-Hubbard 
model in the limit of infinite on-site repulsion (i.e., the case of hardcore bosons), 
in the presence of an alternating on-site chemical potential (a superlattice with period two). 
They showed that this model exhibits all the salient properties apparent
in the Bose-Hubbard model, while also being more `treatable' analytically.
Specifically, they were able to rigorously prove the existence of superfluid and Mott-insulating 
phases in three dimensions. In addition, it is also known that this very same model is exactly solvable in one-dimension
through a mapping to noninteracting fermions. In this case, the half-filled system is 
insulating for any nonzero alternating potential.\cite{Val} The off-diagonal one-particle 
correlations and the momentum distribution function of this model, as well as its nonequilibrium
dynamics, were computed by exact means\cite{rigol04} in Ref.\ \onlinecite{rigol06}.

Motivated by the aforementioned results, here we study the SF-MI phase transition of hardcore 
bosons in the presence of an alternating potential in two and three dimensions. We focus on
the case where the system is at half-filling, in which case 
the transition between the superfluid state and the insulating
state occurs at fixed density. Our first goal is to accurately determine 
the critical values of the alternating potential strength at which the phase transition takes 
place. As a next step, we analyze the results of different approximate solutions,
such as mean-field theory with and without the addition of spin-wave 
corrections, as these allow for an analytical 
treatment of the problem. 

Our approach is to first perform high-precision numerical simulations using the stochastic 
series expansion (SSE) algorithm \cite{SSE1,SSE2} in order to find the critical points of 
the superfluid to Mott-insulator phase transition in the various dimensions. 
The quantities we calculate are the free energy $\Omega$, the density of bosons in the zero-momentum mode $\rho_0$,
\footnote{For homogeneous systems, the number of bosons in the zero-momentum mode coincides with  
the condensate occupation. However, this need not be the case in general, as 
the condensate occupation is defined as the largest eigenvalue of the one-particle density matrix.
The latter quantity will in general be different from the occupation of the zero-momentum 
state if the system is inhomogeneous.\cite{rigol06}}
the superfluid density $\rho_s$ and the compressibility $\kappa=\partial \rho / \partial \mu$. The latter three quantities 
signify the transition from a superfluid to an insulator by dropping to zero at this point (while having 
nonzero values in the superfluid regime). We then employ mean-field and spin-wave
analyses, which allow for some analytical insight into the behavior of our observables of 
interest and the location of the critical point. Our use of these approximation methods 
is partly motivated by results previously reported by Bernardet {\it et al.},\cite{hcb} who studied the 
homogeneous version of the model in two dimensions. There, the mean-field approximation 
alone was shown to provide a fairly good qualitative description of the model,
and remarkably enough, when spin-wave corrections were added, quantities such as 
the superfluid density and the condensate fraction were found to be virtually 
indistinguishable from their exact-numerical counterparts.

The paper is organized as follows. In Sec. \ref{sec:ibh} we briefly review the model at hand.
In Sec. \ref{sec:SSE}, we present the exact numerical solutions obtained using 
the stochastic series expansion (SSE) algorithm. We compute the various physical 
quantities at zero temperature, and find the critical values of the SF-MI phase transition. 
In Sec. \ref{sec:MFSW} we proceed to study several approximation schemes, namely mean-field 
approaches and spin-wave corrections, comparing the critical values obtained using these
methods, with the SSE results. In Sec. \ref{sec:conc} we conclude with a few comments.

\section{\label{sec:ibh}The model}
The Hamiltonian for hardcore bosons in a period-two superlattice in $d$-dimensions, with 
$N=L^d$ sites and periodic boundary conditions, can be written as:
\beq
 \label{eq:Ham}
\hat{H} = - t \sum_{\langle ij \rangle} \left( \hat{a}_i^{\dagger} \hat{a}_j 
+ \hat{a}_j^{\dagger} \hat{a}_i \right) - A \sum_i (-1)^{\sigma(i)} \hat{n}_i -\mu \sum_i \hat{n}_i \,. \nonumber\\
\eeq
Here, $\langle ij \rangle$ denotes nearest neighbors, $\hat{a}_i$ ($\hat{a}_i^{\dagger}$) 
destroys (creates) a hardcore boson on site $i$, $\hat{n}_i=\hat{a}_i^{\dagger} \hat{a}_i$ 
is the local density operator, $\mu$ is the global chemical potential, and 
$ A (-1)^{\sigma(i)}$ is an alternating local potential with $\sigma(i)=0$ on the even 
sublattice and $1$ on the odd sublattice. The hopping parameter $t$ sets the energy scale.

The hardcore boson creation and annihilation operators satisfy the constraints
\begin{equation}
\label{ConstHCB} \hat{a}^{\dagger 2}_{i}= \hat{a}^2_{i}=0, \  
\left\lbrace  \hat{a}_{i},\hat{a}^{\dagger}_{i}\right\rbrace =1, 
\end{equation}
which prohibit double or higher occupancy of lattice sites, as dictated by the 
$U\rightarrow \infty$ limit of the Bose-Hubbard model. 
For any two different sites $i \neq j$, the creation and annihilation operators
obey the usual bosonic relations
\beq
[\hat{a}_{i},\hat{a}_{j}]=[\hat{a}^{\dagger}_{i},\hat{a}^{\dagger}_{j}]=[\hat{a}_{i},\hat{a}^{\dagger}_{j}]=0 \,.
\eeq
\begin{figure}[htp!]
\includegraphics[angle=0,scale=1,width=0.45\textwidth]{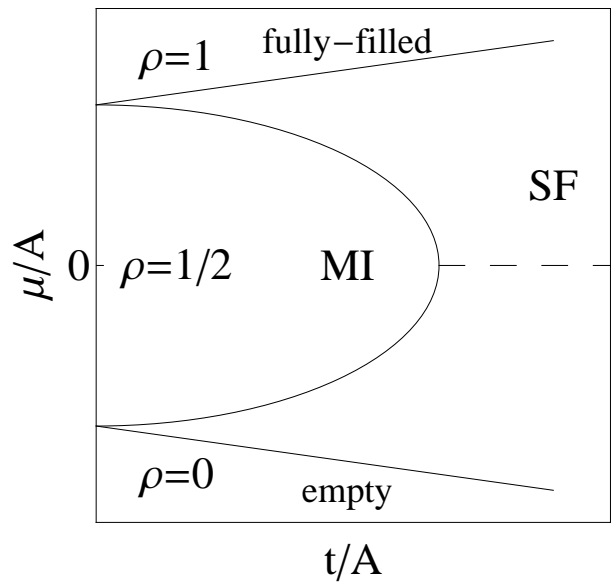}
\caption{\label{fig:lobe} Qualitative description of the expected phase diagram 
of the model at hand, Eq.\ (\ref{eq:Ham}). The diagram contains three insulating
regions corresponding to zero (`empty'), half (`MI') and full (`fully-filled') filling, and a superfluid (SF) phase.}
\end{figure}

The expected phase diagram of the model, in dimensions higher than one, is sketched in 
Fig.\ \ref{fig:lobe}. Our model has two (trivial) insulating regimes corresponding 
to a completely filled lattice (with particle density $\rho=1$), obtained
for large and positive chemical potential values, and a second insulating regime 
which corresponds to an empty lattice, formed in the case where the chemical potential is large and 
negative. These two regimes are also present in the absence of the alternating 
potential. The alternating one-body potential creates another insulating phase, 
one for which the density of particles is $\rho =1/2$. In this case, the alternating 
potential, will in some cases (depending on its strength) create a gap in the energy 
spectrum, generating a superfluid to Mott-insulator transition. 
As the latter regime is the one which is of interest to us, 
we shall henceforth set the global chemical potential to $\mu=0$.
In this case, the model has particle-hole symmetry which in turn fixes the density at $\rho =1/2$
as desired. 

Before moving on, a remark is in order. 
The $\rho=1/2$ insulating phase of the model at hand 
is a consequence of a counterbalance between strong on-site interactions 
(which in our model are in fact infinite) and an alternating potential. 
The resulting local density will thus be different on the odd sublattice than on the 
even sublattice. While this state is sometimes referred to as a charge density wave,\cite{Val}
in what follows, we shall address this phase as a Mott-insulator, in the spirit of Ref.\ \onlinecite{Aizenman}.

\section{\label{sec:SSE}Numerical results}
We obtain numerically-exact results for the model at hand by 
performing numerical simulations based on the stochastic series expansion (SSE) algorithm.\cite{SSE1,SSE2}
As our main objective is to find the critical points of the SF-MI phase transition in the various dimensions, 
simulations are performed for a range of values of the ratio $A/t$ and for various system sizes.
Since we are interested in the zero-temperature properties of the system,
simulations are performed with high inverse-temperature $\beta=1/T$ (in our units, $k_B=1$),
where in most cases we will find it sufficient to have $\beta \geq 2 L$ in order to obtain virtually zero-temperature results.
(The effects of increasing $\beta$ beyond this value are indiscernible.)

Finite size effects are corrected by repeating the simulations with different system sizes.
The thermodynamic-limit value of the phase transition is then extrapolated
by performing finite size scaling of the results in the vicinity of the phase transition:
Around the critical point, most physical quantities (which we denote here by $X$) scale according to the general rule:
\beq \label{eq:scaling}
X L^{\xi/\nu} = F(|A-A_{\textrm{c}}| L^{1/\nu}) \,,
\eeq
where $F$ is a universal scaling function, $A-A_{\textrm{c}}$ is the shifted control parameter 
($A$ being the control parameter, and $A_{\textrm{c}}$ -- the critical value), $\nu$ is the correlation 
length critical exponent and $\xi$ is the critical exponent belonging to the observable $X$.
The values of these exponents are determined by the universality class the transition belongs to.
In our case (and in the Bose-Hubbard model for integer filling as well), it is the ($d+1$) dimensional $XY$ universality 
class.\cite{r4,Bruder} 
We note that the above universality class characterizes only 
the fixed-density transition (the dashed line in Fig.\ \ref{fig:lobe}). The transition driven by
changing the density belongs to the mean-field universality class and is characterized by different critical exponents.

Equation (\ref{eq:scaling}) above will help us find the critical point, 
as it tells us that (a) the quantity  $X L^{\xi/\nu}$ should be independent of the system-size at the phase transition,
and (b) when plotting $X L^{\xi/\nu}$ against $|A-A_{\textrm{c}}| L^{1/\nu}$ the resulting 
curve should be independent of the system-size as well. 

The quantity we shall be using to that end is the superfluid density,
which has the critical exponent $\xi=\nu(d+z-2)$ (see Ref.\ \onlinecite{r4} for details)
where $d$ is the dimension, and $z$ is the dynamical critical exponent, which in our case is $z=1$.\cite{Bruder}
The correlation length exponent $\nu$ is dimension-dependent and takes the values
$1$, $0.672$ and $0.5$ in one, two and three dimensions, respectively. 

\subsection{One dimension}
In one dimension, our model has an analytic solution.\cite{Val}
This is due to the Jordan-Wigner transformation which enables the mapping of the 
hardcore bosons Hamiltonian to that of noninteracting spinless fermions.\cite{Val} 
The latter Hamiltonian may be diagonalized to produce exact analytical results. 
In this case, the SF-MI phase transition occurs at $A_{\textrm{c}}/(2 d t)=0$,
i.e., the system is superfluid only when the alternating potential is absent, in which 
case it exhibits off-diagonal quasi-long-range order (a power-law decay of the one-particle 
correlations). In that sense, one may say that the system exhibits quasi-condensation 
when $A=0$.\cite{Val,rigol04,rigol06} 

\begin{figure}[htp!]
\includegraphics[angle=0,scale=1,width=0.45\textwidth]{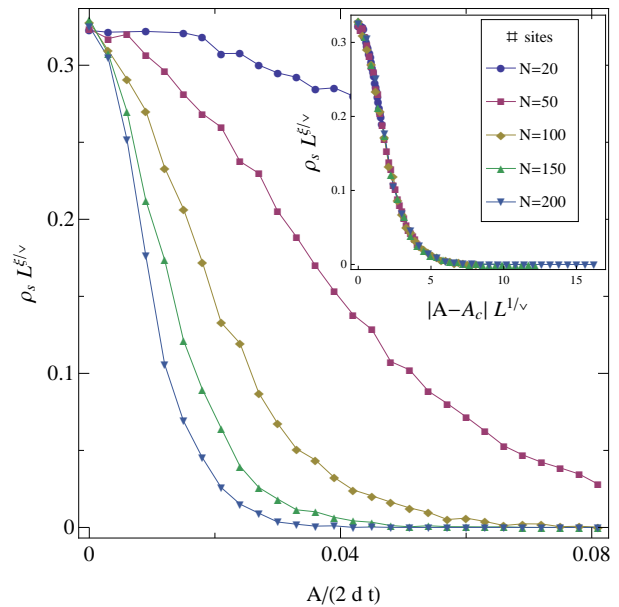}
\caption{\label{dim1} (Color online) Scaled superfluid density as a function of $A/(2 d t)$ for the various system sizes 
in the one-dimensional case. The intersection at $A/(2 d t)=0$ indicates the location
of the SF-MI phase transition. In the inset, the control parameter (the horizontal axis) is scaled as well, 
leading to the collapse of all data points into a single curve.}
\end{figure}

Simulations in one dimension were thus performed only as a check on our computational method. 
No discrepancies between the analytical solution and the numerical one were found:
In Fig.\ \ref{dim1}, the superfluid density is plotted against $A/(2 d t)$ 
for different system sizes (here, $\beta=500$). In the figure, all curves intersect at
the critical point $A_{\textrm{c}}/(2 d t)=0$,
indicating the location of the phase-transition in the thermodynamic limit, in agreement with the analytic results.
The inset shows the scaled superfluid density
as a function of the scaled control parameter, in which case all curves should be, and in fact are, 
on top of each other. The numerical value for the superfluid density at the transition coincides with the expected value of $\pi^{-1}$ given by the analytic solution.\cite{Val} 

\begin{figure}[htp!]
\includegraphics[angle=0,scale=1,width=0.5\textwidth]{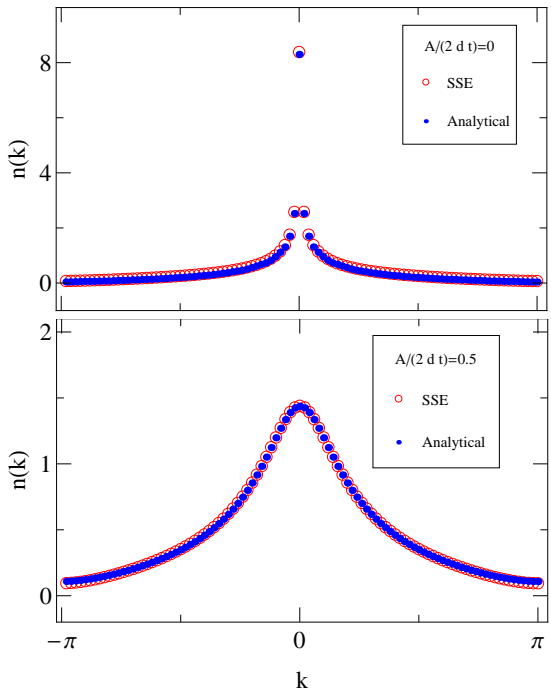}
\caption{\label{mdf1} (Color online) Momentum distribution function $n(k)$ 
in the superfluid regime (top) and in the insulating regime (bottom) for the one-dimensional system with $100$ sites.
In one dimension, the system is superfluid only at $A=0$ (top panel).
This is shown by the sharp peak in the $k=0$ mode of the momentum distribution function which diverges in the thermodynamic limit. 
In this case, the system exhibits quasi-long-range order.   
In both panels, the SSE results (empty circles) are on top of the analytical ones (full circles),
serving as an indication to the accuracy of our computational method.}
\end{figure}

As superlattices such as the one we study here have already been realized 
in experiments with ultracold bosons in optical lattices,\cite{peil03,strabley06,strabley07,lee07} 
and the observable usually measured in those kind of experiments is the momentum distribution function $n(k)$,
we plot it in Fig.\ \ref{mdf1} for two different values of $A/t$. 
Due to the quasi-long-range decay of one-particle correlations in the superfluid phase, the momentum distribution
function has a peak at $k=0$ [Fig.\ \ref{mdf1}(a)]. On the other hand, in the
insulating phase, the one-particle correlations decay exponentially, yielding a broad momentum 
distribution [Fig.\ \ref{mdf1}(b)]. This leads to the following observation: As one increases 
the size of the lattice (while keeping the density fixed), one finds that in the superfluid phase $n(k)$ increases for small values of $k$ 
[Fig.\ \ref{mdf1}(a)], while for the insulating phase [Fig.\ \ref{mdf1}(b)] this does not happen. 
Exact results for $n(k)$ (using the approach described in Ref.\ \onlinecite{rigol04}), are also presented in 
Fig.\ \ref{mdf1}. As expected, the SSE results are right on top of the exact ones. 

\subsection{Two dimensions}
\begin{figure}[htp!]
\includegraphics[angle=0,scale=1,width=0.45\textwidth]{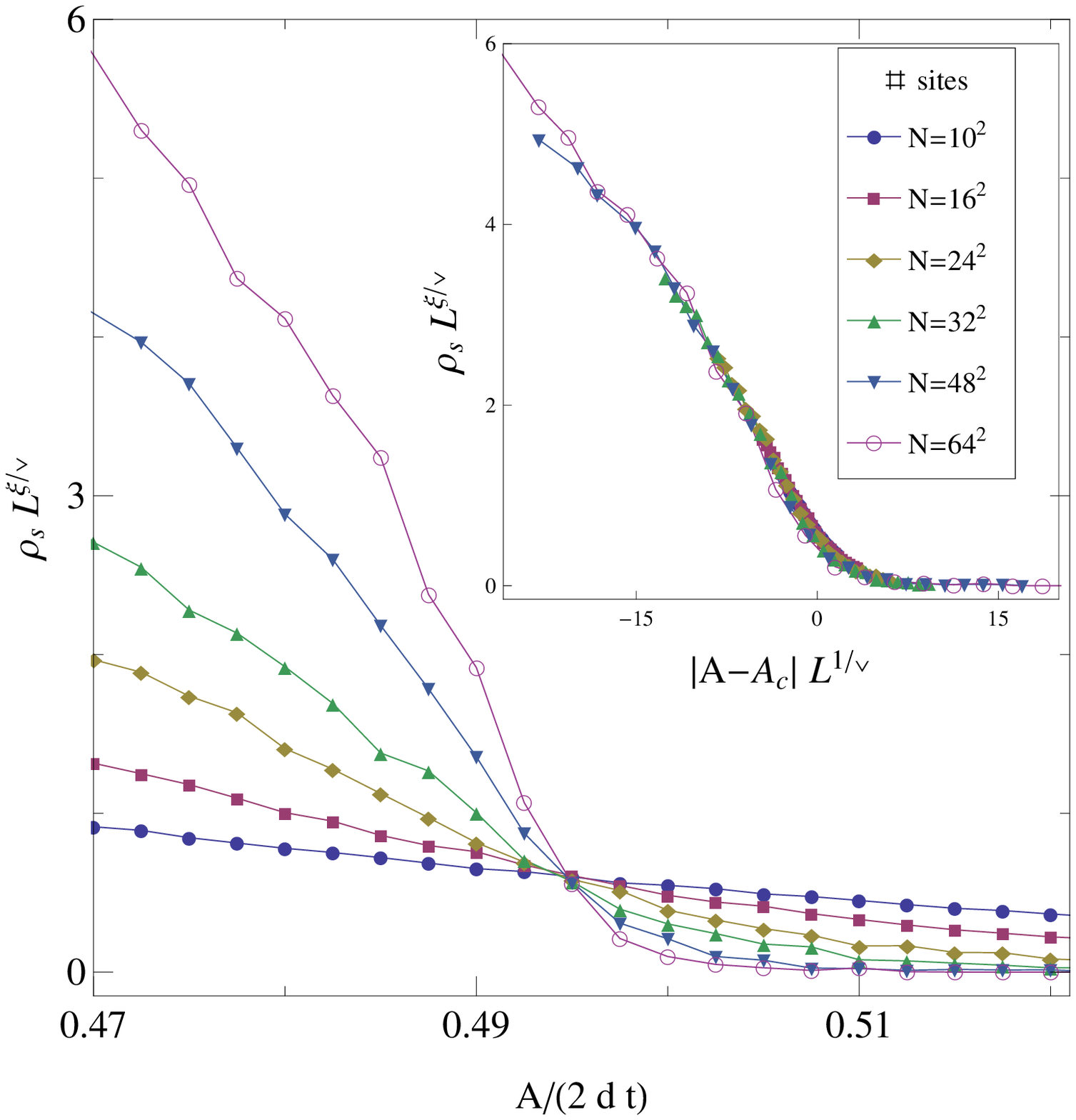}
\caption{\label{dim2} (Color online) Scaled superfluid density as a function of $A/(2 d t)$ for the various 
system sizes in the two-dimensional case. The intersection at $A_{\textrm{c}}/(2 d t) \approx 0.495$ 
indicates the occurrence of the phase transition at that point. In the inset, the control parameter 
(the horizontal axis) is scaled as well, leading to the collapse of all data points into a single curve.}
\end{figure}

In dimensions higher than one, no analytic solution to the model exists, so accurate results are obtainable 
only numerically. Here, we have applied the SSE algorithm to systems of sizes ranging from $10 \times 10$ 
to $64 \times 64$, with inverse-temperature $\beta=96$. In Fig.\ \ref{dim2}, the scaled superfluid density
is plotted against $A/(2 d t)$ for the different system sizes (the errors are on the order of magnitude of the symbol sizes).
All curves intersect at 
$A_{\textrm{c}}/(2 d t) =0.495 (\pm 0.004)$, signifying the phase transition.  
The inset shows the scaled superfluid density as a function of the scaled control parameter.
As in the one-dimensional case, all data points fall into a single curve. 
The value for the critical point we obtained here agrees with 
the value recently obtained by Priyadarshee {\it et al.}\cite{2dcrit}

\begin{figure}[htp!]
\includegraphics[angle=0,scale=1,width=0.45\textwidth]{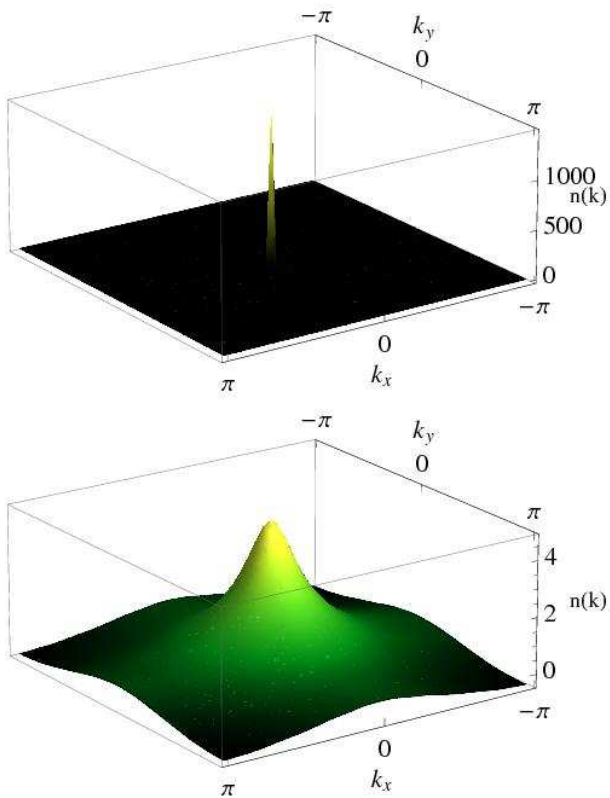}
\caption{\label{mdf2} (Color online) Momentum distribution function $n(k)$ 
in the superfluid regime $A/(2 d t)=0.1$ (top) and in the insulating regime $A/(2 d t)=0.7$ (bottom) 
for a $64 \times 64$ system and $\beta=96$.}
\end{figure}

The momentum distribution function in the superfluid and insulating regimes are shown in 
Figs.\ \ref{mdf2}(a) and \ref{mdf2}(b), respectively. In two dimensions, the superfluid regime exhibits 
true off-diagonal long-range order, which means that the $n(k=0)$ peak is sharper that in one dimension,
which exhibits only quasi-long-range order.
This can be seen in Fig.\ \ref{mdf2}(a). The Mott-insulating regime is once again characterized by 
an exponential decay of one-particle correlations. The corresponding momentum distribution 
function has a broad peak around $n(k=0)$ as shown in Fig.\ \ref{mdf2}(b). 

\subsection{Three dimensions}
\begin{figure}[htp!]
\includegraphics[angle=0,scale=1,width=0.45\textwidth]{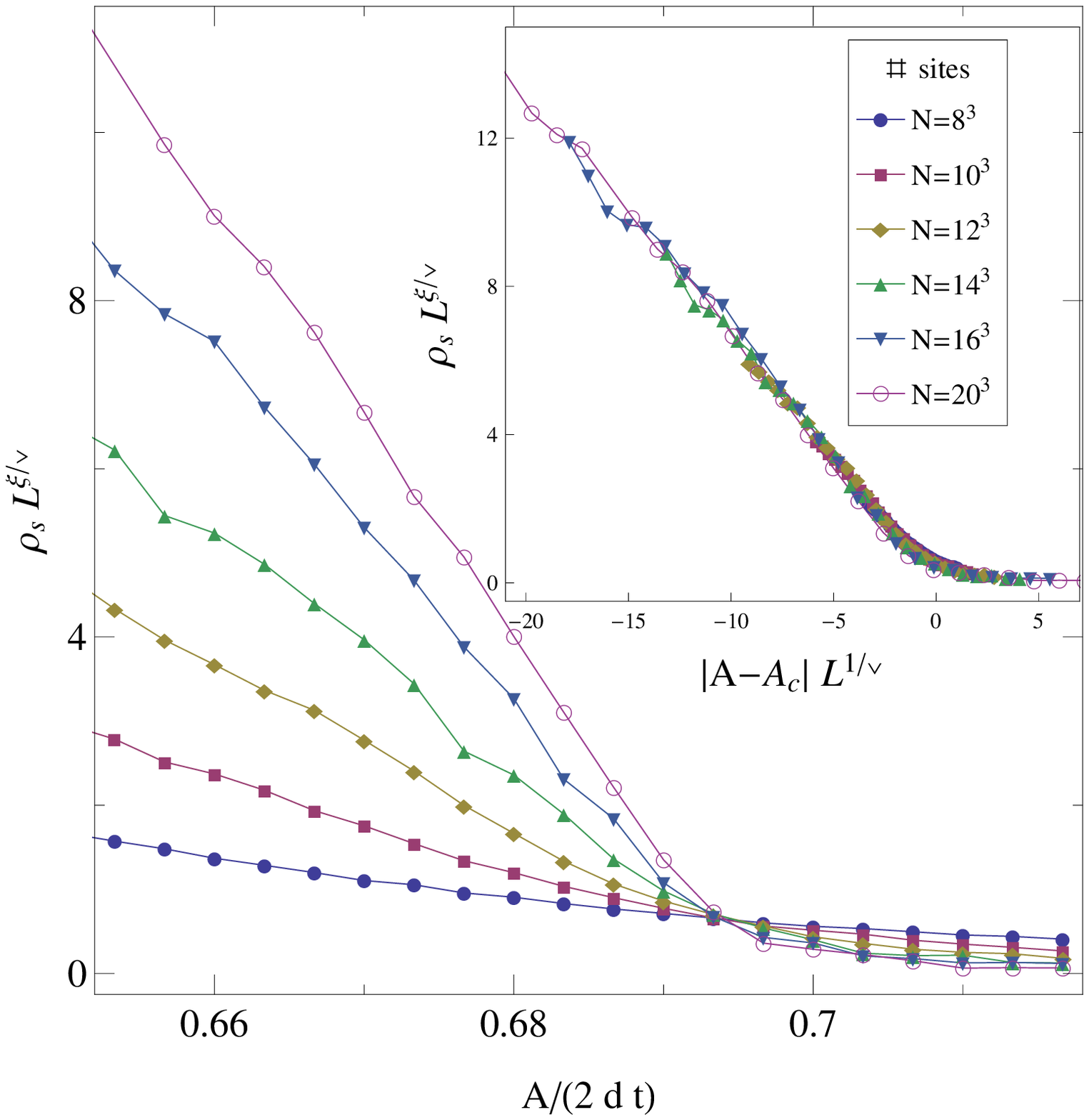}
\caption{\label{dim3} (Color online) Scaled superfluid density as a function of $A/(2 d t)$ for the various system sizes 
in the three-dimensional case. The intersection at $A_{\textrm{c}} /(2 d t) \approx 0.693$
indicates the location of the SF-MI phase transition.
In the inset, the control parameter (the horizontal axis) is scaled as well, 
leading to the collapse of all data points into a single curve.}
\end{figure}

In three dimensions, we have performed simulations for system sizes ranging from 
$6 \times 6 \times 6$ to $20 \times 20 \times 20$ and an inverse temperature of $\beta=40$. 
Here, the SF-MI phase transition is found at $A_{\textrm{c}} /(2 d t) = 0.693 (\pm 0.005)$,
as indicated by the scaled superfluid density plotted as a function of $A /(2 d t)$ in 
Fig.\ \ref{dim3}, for the different system sizes. The inset in Fig.\ \ref{dim3} depicts 
the scaled superfluid density as a function of the scaled control parameter, exhibiting 
the collapse of all data points into a single curve, as in one and two dimensions. 
The momentum distribution function in three dimensions is qualitatively similar to
its two-dimensional counterpart, both in the superfluid phase and in the insulating phase, and thus will not be
presented here.

\section{\label{sec:MFSW}Approximation schemes}
Having obtained the critical values via quantum Monte Carlo techniques, 
we now turn to look for approximation schemes that would provide analytical descriptions 
of the phase transition. We start this investigation with the Gutzwiller mean-field 
approach. Before doing so however, we recall that the model at hand can also be viewed 
as the $XY$ model of a spin-1/2 system.\cite{mats} We shall make use of this correspondence, 
utilizing the exact mapping between bosonic operators and $SU(2)$ generators, namely,
\beq
\hat{a}_i^{\dagger} &\leftrightarrow & S_i^{+} \,, \\\nonumber
\hat{a}_i &\leftrightarrow & S_i^{-} \,, \\\nonumber
\hat{a}_i^{\dagger} \hat{a}_i &\leftrightarrow & S_i^{z}+1/2 \,.
\eeq
With this mapping, the hardcore bosons Hamiltonian, \hbox{Eq.\ (\ref{eq:Ham})}, becomes
that of the $XY$ antiferromagnet with an alternating magnetic field applied along the $\hat{z}$ direction:
\beq
\hat{H}= &-& t \sum_{\langle ij \rangle} \left( S_i^{+} S_j^{-} + S_j^{+} S_i^{-} \right) 
\nonumber \\ &-& \sum_i \left[ \mu +A (-1)^{\sigma(i)} \right] \left( S_i^{z}+\frac1{2} \right)  \,.
\eeq

\subsection{\label{sec:mf}Mean-field approach}
\begin{figure}[hbp!]
\includegraphics[angle=0,scale=1,width=0.25\textwidth]{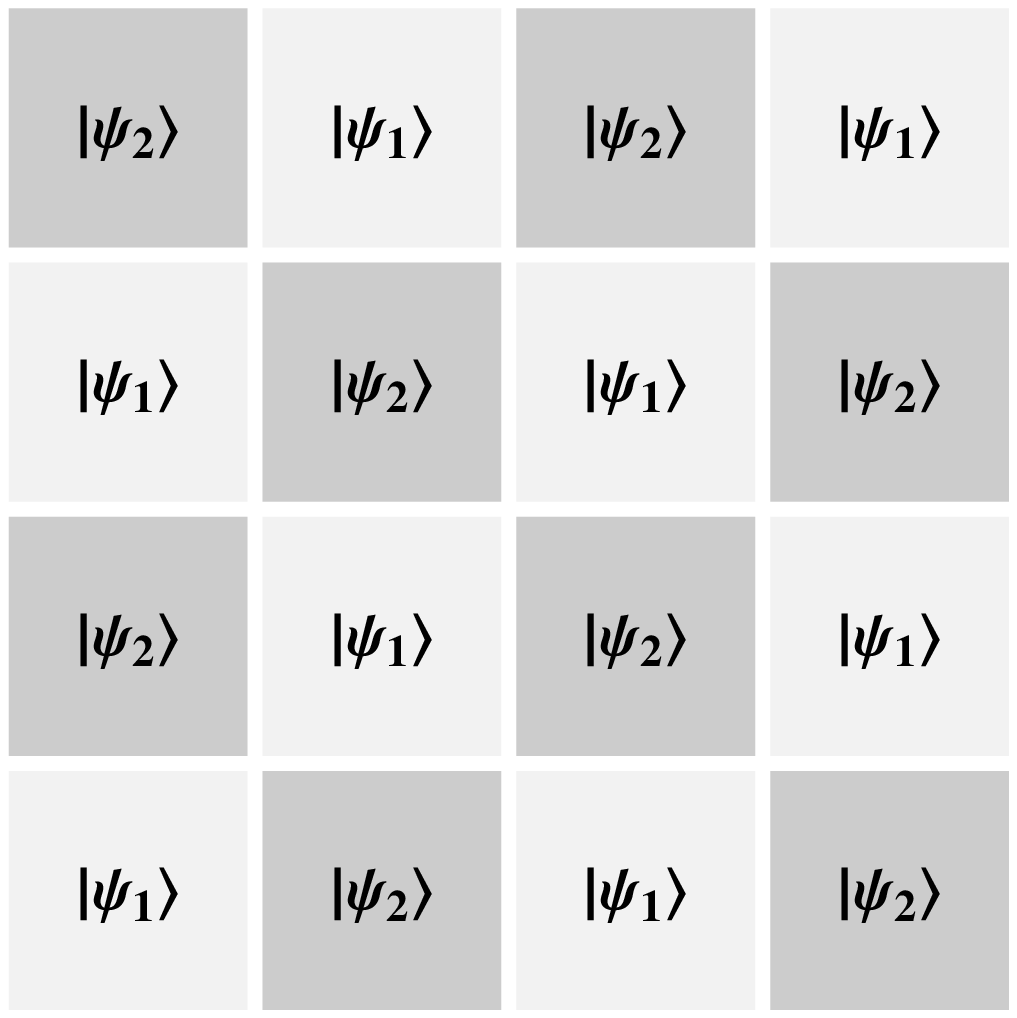}
\caption{\label{Figure7} A schematic description of the product state in the mean-field
approach in two dimensions. Every other site is described by the 
same wave function.}
\end{figure}

We start our mean-field calculation with the following product state as an initial ansatz:
\beq \label{eq:GSMF}
| \GS \rangle_{\mf} =\prod_j^{\otimes} \left( \sin \frac{\theta_j}{2} | \downarrow \rangle + 
\cos \frac{\theta_j}{2} \e^{i \, \varphi_j} | \uparrow \rangle 
\right) \,,
\eeq
where $(\theta_j,\varphi_j)$ specify the orientation of the $j$-th spin.
Obviously, we expect every other site to be described by the same wave function,
due to the symmetry of the problem. 
This is schematically shown in Fig.\ \ref{Figure7}.
As we are using the grand-canonical scheme, the orientations of the spins  will be determined
by minimizing the grand-canonical potential (per site)
\beq \label{eq:omegaMF}
\Omega_{\mf} &=& {_{\mf}}\langle \GS| \hat{H} | \GS\rangle_{\mf} =-\frac{t}{2 N} \sum_{\langle ij \rangle} \sin \theta_i \sin \theta_j \cos(\phi_i-\phi_j)
\nonumber\\
&\phantom{=}&-\frac1{2 N} \sum_i \left[ \mu +A (-1)^{\sigma(i)} \right] \left(1+\cos \theta_i\right)\,.
\eeq
with respect to these angles. 
For the azimuthal angles, this simply implies a constant (yet arbitrary) value \hbox{$\varphi_j=\Phi$}, while
for the polar angles, extremization yields
\begin{subequations} \label{eq:thetas}
\begin{align}
\cos \theta_1 &= \mu_1 \sqrt{\frac{1+{\mu_2}^2}{1+{\mu_1}^2}} \,,\\
\cos \theta_2 &= {\mu_2} \sqrt{ \frac{1+{\mu_1}^2}{1+{\mu_2}^2}} \,,
\end{align}
\end{subequations}
where $\mu_{1,2} \equiv (\mu \pm A)/(2 d t)$.
These values correspond to a minimal configuration only in the region
$|\mu_1 \mu_2|<1$. Outside this region, the system is saturated, and 
the solution is one where all spins are aligned -- pointing either all up or all down. 
In bosonic language, these latter configurations correspond to the completely full/empty insulating
configurations.
\par
At this point we can calculate the following quantities. 
First, the density of particles is:
\beq
\rho_{\mf} &=&\frac1{N} \sum_i {_\mf}\langle \GS | \hat{a}_i^{\dagger} \hat{a}_i | \GS \rangle_{\mf} =
\frac1{2}+ \frac1{2 N}\sum_i \cos \theta_i
\nonumber\\
&=&\frac1{2} + \frac1{4} \left( \cos \theta_1 + \cos \theta_2\right)
\,.
\eeq
Next, the free energy becomes
\beq
\Omega_{\mf} &=& {_{\mf}}\langle \GS| \hat{H} | \GS\rangle_{\mf}
=- \frac{d t}{2} \sin \theta_1 \sin \theta_2 - \frac{\mu}{2} \nonumber\\
&-&\frac1{4} \left( \mu +A \right) \cos \theta_1
-\frac1{4} \left( \mu -A \right) \cos \theta_2 \,,
\eeq
and the density of bosons in the zero-momentum mode $\rho_0$ is calculated as:
\beq
\rho_{0,\mf} &=& \frac1{N} {_{\mf}}\langle \GS |\hat{a}^{\dagger}_{\bk=0} \hat{a}_{\bk=0}  | \GS \rangle_{\mf}
\\\nonumber
&=&\frac1{4 N^2} \sum_{i,j} \sin \theta_i \sin \theta_j =\frac1{16} \left( \sin \theta_1 + \sin \theta_2 \right)^2 \,.
\eeq
\par
The superfluid density $\rho_s$ requires a special treatment of the boundary conditions. 
As is well known,\cite{rhoS} one can relate the superfluid density to the ``spin stiffness''. To accomplish this,
one needs to compare $\Omega$ (the free energy) of the system under periodic conditions with the free energy 
under a ``twist'' in the boundary conditions along one of the linear directions (say, the $x$ direction). 
In the periodic case, which we treated above, the azimuthal angles $\varphi_j$ were all identical.
To implement a twist, we take this angle to be site-dependent and with a constant gradient
such that the total twist across the system in the $x$ direction is $\pi$, namely 
$\delta \varphi = \varphi_{j+\hat{x}}-\varphi_{j}= \pi/L$. 
Within the mean-field treatment, one can show that addition of this gradient 
is equivalent to substituting $t \to t/d \left[ (d-1) + \cos \delta \varphi\right]$.
Now, the square of the gradient twist is related to the superfluid density via,\cite{rhoS,hcb}
\beq
\Omega_{\textrm{twisted}}-\Omega= t \rho_s  \delta \varphi^2 \,,
\eeq
which in turn yields the simple expression
\beq \label{eq:rhos}
\rho_s = -\frac1{2d} \frac{\partial \Omega}{\partial t} \,.
\eeq
\par
\begin{figure}[htp!]
\includegraphics[angle=0,scale=1,width=0.5\textwidth]{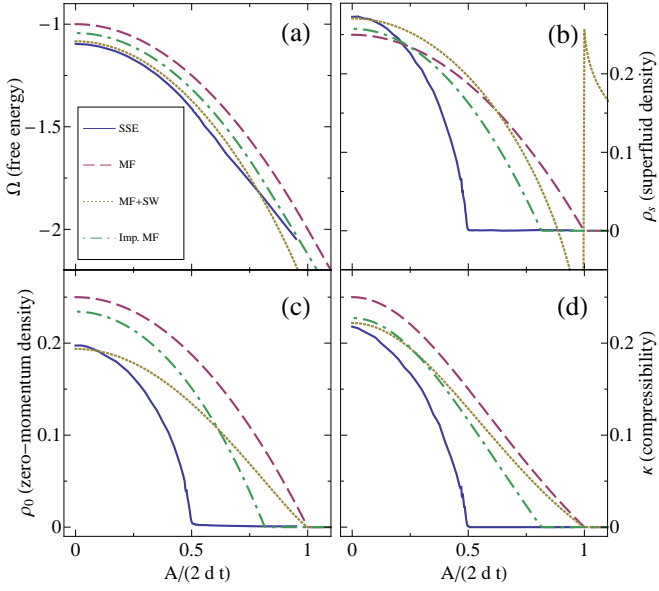}
\caption{\label{Figure8} (Color online) Thermodynamic quantities in two dimensions. 
(a) Free energy [$t=1$], (b) superfluid density, (c) the density of bosons in the zero-momentum mode, and (d) compressibility 
as a function of $A/(2 d t)$. The solid lines indicate the SSE results 
($64 \times 64$ sites, $\beta=96$), whereas the dashed, dotted and dash-dotted lines 
indicate the mean-field, mean-field plus spin-waves and improved mean-field results, 
respectively.}
\end{figure}
\begin{figure}[htp!]
\includegraphics[angle=0,scale=1,width=0.5\textwidth]{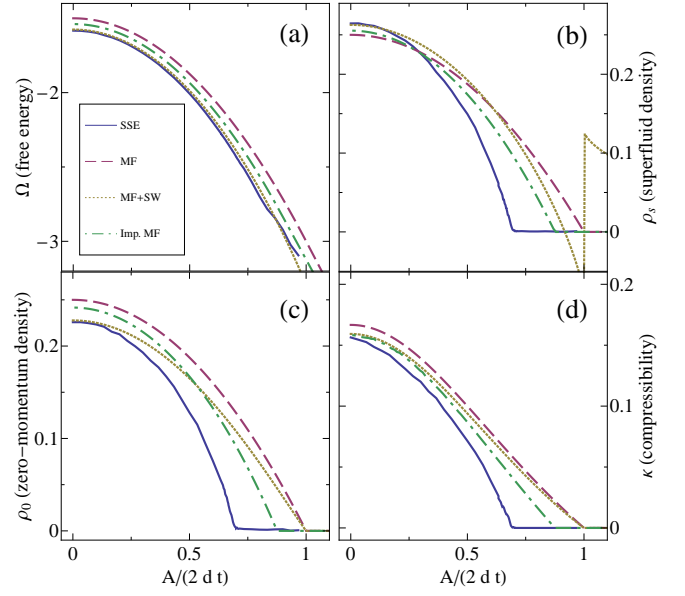}
\caption{\label{Figure9} (Color online) Thermodynamic quantities in three dimensions.
(a) Free energy [$t=1$], (b) superfluid density, (c) the density of bosons in the zero-momentum mode, and (d) compressibility 
as a function of $A/(2 d t)$. The solid lines indicate the SSE results 
($16 \times 16 \times 16$ sites, $\beta=40$), whereas the dashed, dotted and dash-dotted 
lines indicate the mean-field, mean-field plus spin-waves and improved mean-field results, 
respectively.}
\end{figure}

Setting $\mu=0$, this expression for the superfluid density coincides with that of $\rho_{0,\mf}$:
\beq
\rho_{s,\mf}=\rho_{0,\mf} = \Bigg\{
\begin{tabular}{ccc}
$\frac1{4} - \left( \frac{A}{4 d t} \right)^2$ &,& $\frac{A}{2dt}<1$\\ 
$0$ &,&  $\frac{A}{2dt}\geq 1$ 
\end{tabular} \,,
\eeq
giving the critical value for the phase transition $A_{\textrm{c}}/(2 d t)=1$. 
Figures \ref{Figure8} and \ref{Figure9} show: (a) the free energy, (b) the superfluid density,
(c) the density of bosons in the zero-momentum mode, and (d) the compressibility of the system as a function of $A /(2 d t)$
in two and three dimensions. The dashed and solid curves represent the mean-field and SSE results, 
respectively. As one can immediately see, the critical values obtained within the mean-field 
approximation do not agree with the exact-numerical results. In two dimensions the error is 
$\approx 100\%$ and in three dimensions, it is $\approx 50\%$.
The very large errors here merely reflect the fact that the mean-field approach used here is not 
fit to describe the model at hand, especially in the vicinity of the SF-MI phase transition. 

\par
\subsection{Adding spin-wave corrections}
As pointed out earlier, the addition of spin-wave corrections yields virtually exact 
results in the homogeneous case in two dimensions.\cite{hcb} For the reader's convenience, 
we review the mean-field calculations of the homogeneous ($A=0$) case and its spin-wave 
corrections in Appendix\ \ref{app:mf} (thereby also correcting some misprints that appeared in
the original manuscript examining this case, Ref.\ \onlinecite{hcb}). Let us see how the mean-field results are modified by 
the addition of spin-wave corrections in our case. 
To include these, we proceed in the usual way.\cite{SW1,SW2,SW3,SW4}
We first introduce a set of local rotations that align the $\hat{z}$ direction 
of each of the spins with its mean field orientation.
This is accomplished by switching to new spin operators defined by
\beq
\left( 
\begin{tabular}{c}
$S'^x_j$\\
$S'^y_j$\\
$S'^z_j$\\
\end{tabular}
\right)=
R(\theta_j,\varphi_j)
\left( 
\begin{tabular}{c}
$S^x_j$\\
$S^y_j$\\
$S^z_j$\\
\end{tabular}
\right)
\eeq
where $R(\theta_j,\varphi_j)$ is the $3 \times 3$ rotation matrix
\beq
R(\theta_j,\varphi_j) = \left(
\begin{tabular}{ccc}
$\cos \theta_j \cos \varphi_j$ & $-\sin \varphi_j$ & $\sin \theta_j \cos \varphi_j$  \\
$\cos \theta_j \sin \varphi_j$ & $\cos \varphi_j$ & $\sin \theta_j \sin \varphi_j$\\
$-\sin \theta_j $ & 0 & $\cos \theta_j$ \\
\end{tabular}
\right) \,.
\eeq
The corresponding new annihilation and creation operators $\hat{b}_j \leftrightarrow S'^{-}_j$ 
and $\hat{b}_j^{\dagger} \leftrightarrow S'^{+}_j$ 
describe low-energy
fluctuations about the mean-field ground state -- 
these are spin waves. They too obey hardcore bosons commutation relations. 
Substituting these expressions into our Hamiltonian, and ignoring cubic and quartic terms 
in these bosonic operators (thus assuming a dilute gas of spin waves), the new Hamiltonian 
reads
\beq \label{eq:Hsw}
\hat{H}_{\sw} &=& \hat{H}_{\mf} + D  \sum_i \hat{b}_i^{\dagger} \hat{b}_i + C  \sum_i (-1)^{\sigma(i)} \hat{b}_i^{\dagger} \hat{b}_i 
\\\nonumber
&+& B \sum_{\langle i j \rangle} (\hat{b}_i^{\dagger} \hat{b}_j^{\dagger} + \hat{b}_i \hat{b}_j )
  - \frac{A}{2} \sum_{\langle i j \rangle} (\hat{b}_i^{\dagger} \hat{b}_j + \hat{b}_i \hat{b}_j^{\dagger}  ) \,,
\eeq
where the coefficients are 
\begin{subequations}
\begin{align}
A &= t \left(  1+\cos \theta_1 \cos \theta_2 \right) \,, \\
B &= t/2 \left( 1-\cos \theta_1 \cos \theta_2 \right)\,, \\
C &= d t \left( \mu_1 \cos \theta_1 - \mu_2 \cos \theta_2 \right) \,,\\
D &= d t \left( 2 \sin \theta_1 \sin \theta_2 +\mu_1 \cos \theta_1 + \mu_2 \cos \theta_2 \right) \,.
\end{align}
\end{subequations}
This quadratic Hamiltonian can be diagonalized by first going to Fourier space,
using $\hat{b}_{j} = N^{-1/2} \sum_{\bk} \e^{2 \pi i \bk j /L} \hat{b}_{\bk}$.
This in turn yields the Hamiltonian:
\beq \label{eq:Hsw2}
\hat{H}_{\sw} = \hat{H}_{\mf} &+& \sum_{ \bk}  (D-A \gamma_{\bk}) \hat{b}_{\bk}^{\dagger} \hat{b}_{\bk} 
+ C\sum_{ \bk}  \hat{b}_{\bk}^{\dagger} \hat{b}_{\bk+L/2}\nonumber\\
&+& B \sum_{ \bk}   \gamma_{\bk} (\hat{b}_{\bk}^{\dagger} \hat{b}_{L-k}^{\dagger}+\hat{b}_{\bk} \hat{b}_{L-{\bk}}) \,,
\eeq
where, $\gamma_{\bk}=\sum_{i=1}^{d} \cos \left( \frac{2 \pi k_i}{L} \right)$, 
and $k_1 \ldots k_d$ are the components of the momentum vector in each of the directions.
We note that the Fourier-space operators $\hat{b}_{\bk}$ and $\hat{b}_{\bk}^{\dagger}$ 
no longer obey the hardcore bosons commutation relations. These field operators are only excitations 
about the ground state, and are treated as soft-core bosons.\cite{hcb,SW1,SW2,SW3,SW4}
At this point, our Hamiltonian may be diagonalized in a straightforward manner 
(we review the diagonalization process in Appendix\ \ref{app:diag}).
Once diagonalized, the Hamiltonian takes the form
\beq \label{eq:hsw}
\hat{H}_{\sw} = \hat{H}_{\mf} + \sum_{\bk} \Lambda_{\bk} \hat{\eta}_{\bk}^{\dagger} \hat{\eta}_{\bk} +E_0 \,,
\eeq
where the $\Lambda_{\bk}$'s are energy levels and $E_0$ is 
the correction to the ground-state energy of the system, given by:
\begin{widetext}
\beq
E_0=\frac1{4} \sum_k \Bigg(  -2D 
&+&\sqrt{(A^2-4B^2) \gamma_k^2 +D^2 +C^2 +2\sqrt{(D C)^2 +[(A D)^2 -(2 B C)^2] \gamma_k^2}} \nonumber\\
&+&\sqrt{(A^2-4B^2) \gamma_k^2 +D^2 +C^2 -2\sqrt{(D C)^2 +[(A D)^2 -(2 B C)^2] \gamma_k^2}} \,\Bigg) \,.
\eeq
\end{widetext}
The operators $\hat{\eta}_{\bk}^{\dagger}$ and $\hat{\eta}_{\bk}$ in Eq.\ (\ref{eq:hsw}) are modified 
spin-wave creation and annihilation operators, respectively,
and are each a linear combination of $\hat{b}_{\bk}, \hat{b}_{L-\bk}, \hat{b}_{\bk+L/2}, \hat{b}_{L/2-k}$ 
and their adjoints. The coefficients of these linear combinations are fixed during the diagonalization process,
and using them, all physical observables can be calculated in a straightforward manner 
(we elaborate on this matter in Appendix\ \ref{app:diag}).   

The results of the spin-wave analysis are indicated by the dotted lines in Figs.\ \ref{Figure8} 
(two dimensions) and \ref{Figure9} (three dimensions). They show: (a) the free energy, (b) the superfluid density,
(c) the density of bosons in the zero-momentum mode, and (d) the compressibility, after the addition of spin-wave 
corrections, as a function of $A /(2 d t)$. 

As one can see in those figures, in the superfluid phase, the spin-wave corrected 
values for the free energy are almost on top of the exact-numerical ones;
and more so in the three-dimensional case than in the two-dimensional one. 
As for the other measured observables, the spin-wave corrections are clearly 
an improvement over the mean-field results, especially for small values of $A/t$ 
where the spin-wave corrections yield virtually exact results.
Unfortunately however, as one approaches the phase transition itself, 
the spin-wave corrections lose their accuracy, eventually leaving the 
phase-transition at its mean-field value, namely at $A_{\textrm{c}}/(2 d t)=1$. 

Another issue worth noting here is the behavior of the spin-wave corrected 
superfluid density [Figs.\ \ref{Figure8}(b) and\ \ref{Figure9}(b)] in the 
vicinity of the predicted phase transition, $A/(2 d t)=1$. 
On the superfluid side of the transition the superfluid density becomes 
negative, indicating the breakdown of the spin-wave approximation for that 
quantity. The transition point is still signaled by a discontinuity in 
$\rho_s$. However, the overall behavior of the superfluid density around the transition point is 
clearly an artifact of the spin-wave approximation and should not be considered further.

\subsection{Improved mean-field approach}
Having seen that spin-wave corrections, albeit accurate in the weak-potential regime,
do not modify the critical 
point predicted by the mean-field solution, we have devised an improved mean-field approach. 
As we show now, this method provides a significant improvement over the mean-field results (and the spin-wave corrections) 
discussed previously, particularly in the context of the location of the critical point.

\begin{figure}[htp!]
\includegraphics[angle=0,scale=1,width=0.25\textwidth]{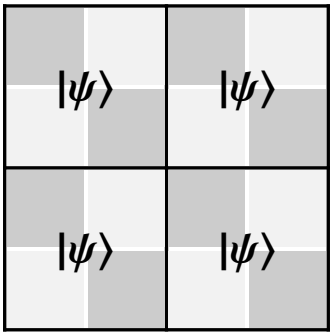}
\caption{\label{Figure10} In the `improved mean-field' case, 
a larger unit-cell is defined. In the two-dimensional case at hand, the new cell consists of $2 \times 2$ sites.
With this new definition the model turns homogeneous and a product of identical wave functions is then guessed
as a solution.}
\end{figure}

We start with a variational ansatz which, as before, is a product state. 
However, this time we do not choose a product of single-site wave-functions. The new ansatz is a 
product of wave-functions each describing the state of a `block' of $2^d$ sites, such that with 
this block as the basic cell, the model turns homogeneous. In two dimensions a block consists 
of $2 \times 2$ cells (as shown in Fig.\ \ref{Figure10}) each of which is described 
by the general wave function
\beq \label{eq:gsimf}
| \GS \rangle_{\imf} =\prod_{\textrm{blocks}}^{\otimes} 
\left(\sum_{i,j,k,l \in \{ \downarrow,\uparrow \}} c_{ijkl} | i j k l\rangle \right) \,,
\eeq
where the generalization to three dimensions, in which case the basic block is a $2 \times 2 \times 2$ cell,
is straightforward (note that the coefficients for each of the blocks are the same).
As before, we minimize the free energy 
\hbox{$\Omega_{\imf} = {_{\imf}}\langle \GS | \hat{H} | \GS \rangle_{\imf}$} with respect 
to the coefficients $c_{ijkl}$ of the wave function (this time we do so numerically).
Obtaining the various observables in terms of the wave function  given in Eq.\ (\ref{eq:gsimf}) is straightforward,
and was performed in much the same way as the usual mean-field approach discussed in Sec.\ \ref{sec:mf}. 
The results of this approximation are given by the 
dash-dotted lines in Figs.\ \ref{Figure8} and \ref{Figure9}. They
depict: (a) the free energy, (b) the superfluid density,
(c) the density of bosons in the zero-momentum mode, and (d) the compressibility, as a function of $A/(2 d t)$.

As the figures indicate, in most instances, the results of this method are more accurate 
than those of the previous approximation schemes, in particular, for the location of 
the phase transition. The critical values given by this approximation are 
$A_{\textrm{c}}/(2 d t)=0.815$ in two dimensions ($\approx 60\%$ error) 
and $A_{\textrm{c}}/(2 d t)=0.875$ in three dimensions ($\approx 24\%$ error). 
Also, we note that while the spin-wave corrected values for the various 
thermodynamic quantities are a better approximation 
in the weak potential [small $A/(2 d t)$] regime, as one moves away from
this region, the improved mean-field technique proves to be a better estimator for 
all quantities but the free energy. It is clear still that the improved-mean-field method presented here 
is far from being very accurate.

\section{\label{sec:conc}Conclusions}

We have studied the superfluid to Mott-insulator phase transition of hardcore 
bosons in a period-two superlattice in two and three dimensions. We focused on the case 
where the system is at half filling, for which the quantum phase transition belongs to 
the $(d+1)$ dimensional $XY$ universality class. 

Using quantum Monte Carlo simulations and finite size scaling, we have determined the critical 
value of the alternating potential parameter $A$ at which the SF-MI phase transition occurs. 
In two dimensions, our results agree with previous calculations.\cite{2dcrit}

We have also compared our numerical results against several approximation schemes, some of which have been used 
successfully in the two-dimensional homogeneous version of the model. 
We have seen that employing a mean-field approach using the usual Gutzwiller ansatz works very poorly 
($\approx 100\%$ error in two dimensions and $\approx 50\%$ error in three dimensions).
This is a clear indication of the fact that this mean-field approach
is not suitable for describing this model, especially in the vicinity of the phase transition,
as it breaks down in the strong coupling regime. 

The spin-wave corrections to the mean-field solution turned out to be very useful, especially in
the superfluid phase, where the spin-wave corrected estimation of the free-energy is very close to 
the exact values, and also reproduced the exact results for all observables for small values 
of $A/t$. However, as one moves away from the weak potential regime, the spin-wave corrections become 
more and more inaccurate, and their predictions of the critical points eventually coincide with 
those of the mean-field approach, therefore indicating their unusefulness in that region.

The improved mean-field approximation scheme we have devised here, which was
based on the underlying homogeneity of the problem, has proved to be 
an improvement over the previous methods, albeit still far from being accurate. 
This approach provides an analytical description of the superfluid-to-Mott-insulator 
transition and gives an estimate of the critical value for the transition with (around) 
one half the error of the usual Gutzwiller ansatz, i.e., it is an improvement 
in terms of the location of the critical point.

\begin{acknowledgments}
This work was supported by the US Office of Naval Research
Award No.\ N000140910966 and by startup funds from Georgetown University. 
We thank Tommaso Roscilde and Valery G. Rousseau for useful discussions.
\end{acknowledgments}

\appendix

\section{\label{app:mf}Mean-field results and spin-wave corrections in the homogeneous case}
In what follows, we briefly review the results of the mean-field approximation of Sec.\ \ref{sec:mf} 
and its spin-wave corrections in the homogeneous ($A=0$) case for arbitrary values of $\mu$.

Starting with the ansatz given in Eq.\ (\ref{eq:GSMF}), minimization of the 
free energy Eq.\ (\ref{eq:omegaMF}) with respect to the spin orientation angles yields 
\beq
\cos \theta_j = \frac{\mu}{2 d t} \,, 
\eeq 
where the azimuthal angle takes on, once again, a constant yet arbitrary value \hbox{$\varphi_j=\Phi$}.
The density of particles becomes
\beq
\rho_{\mf}&=&\frac1{N} \sum_i {_{\mf}}\langle \GS| \hat{a}_i^{\dagger} \hat{a}_i | \GS \rangle_{\mf} =
\frac1{2}+ \frac1{2 N}\sum_i \cos \theta_i
\nonumber\\
&=&\frac1{2}\left(1+\frac{\mu}{2 d t}\right)\,,
\eeq
and the free energy is
\beq
\Omega_{\mf} &=&{_{\mf}}\langle \GS| \hat{H} | \GS\rangle_{\mf} =- \frac{d t}{2} \sin^2 \theta -\frac1{2} \mu(1+ \cos \theta)
\nonumber\\
&=& -\frac1{2} d t \left( 1+\frac{\mu}{2 d t}\right) ^2 \,.
\eeq
The density of bosons in the zero-momentum mode turns out to be
\beq
\rho_{0,\mf}&=& \frac1{N} {_{\mf}}\langle \GS |\hat{a}^{\dagger}_{\bk=0} \hat{a}_{\bk=0}  | \GS \rangle_{\mf}
\\\nonumber
&=&\frac1{4 N^2} \sum_{i,j} \sin \theta_i \sin \theta_j =\frac1{4} \left[1-\left(\frac{\mu}{2 d t} \right)^2 \right] \,.
\eeq
Using Eq.\ (\ref{eq:rhos}), it can be easily shown that the expression for the superfluid density $\rho_{s,\mf}$ 
in the homogeneous case coincides with the expression obtained for $\rho_{0,\mf}$ above (as is the case
with the alternating potential). 

The addition of spin-wave corrections to the mean-field results 
is carried out in exactly the same manner as with the staggered potential.
The Hamiltonian in this case has the same form as the one given in Eq.\ (\ref{eq:Hsw}) but with coefficients
\begin{subequations}
\begin{align}
A &= t \left[ 1+\left( \frac{\mu}{2 d t} \right)^2 \right] \,, \\
B &= \frac{t}{2} \left[ 1 - \left( \frac{\mu}{2 d t} \right)^2 \right] \,, \\
C &= 0 \,,\\
D &= 2 d t \,.
\end{align}
\end{subequations}
The spin-wave field operators which diagonalize the 
Hamiltonian are given by the simple relation:\cite{hcb}
\beq
\hat{b}_k= \cosh \phi_k \, \hat{\eta}_k - \sinh \phi_k \, \hat{\eta}^{\dagger}_{L-k}\,,
\eeq
with $\phi_k$ obeying
\begin{subequations}
\begin{align}
\sinh^2 \phi_k &= \frac1{2} \left( \frac{D-A \gamma_k}{\sqrt{(D-A \gamma_k)^2-(2 B \gamma_k)^2}} -1 \right) \,,\\
\cosh^2 \phi_k &= \frac1{2} \left( \frac{D-A \gamma_k}{\sqrt{(D-A \gamma_k)^2-(2 B \gamma_k)^2}} +1 \right) \,,
\end{align}
\end{subequations}
and so, the various spin-wave corrected quantities may be written explicitly: 
the corrected density of particles is
\beq
\rho_{\sw}= \rho_{\mf} -\frac1{N} \frac{\mu}{2 d t} \sum_{ k \neq 0} \sinh^2 \phi_k\,,
\eeq
and the free energy becomes
\beq
\Omega_{\sw}&=& \Omega_{\mf}\\ \nonumber
&+& \frac1{2} \sum_{k \neq 0} \left[ \sqrt{(D-A \gamma_k)^2-(2 B \gamma_k)^2} -(D-A \gamma_k) \right] \,.
\eeq
Using Eq.\ (\ref{eq:rhos}), the superfluid density immediately follows. 
Finally, the density of bosons in the zero-momentum mode turns out to be:
\beq
\rho_{0,\sw}=\rho_{0,\mf} -\frac1{N} \left[ 1- \left( \frac{\mu}{2 d t} \right)^2 \right]  \sum_{k \neq 0} \sinh^2 \phi_k
\,.\nonumber\\
\eeq

\section{\label{app:diag} Diagonalization of quadratic bosonic Hamiltonians}
Following the prescription given in Ref.\ \onlinecite{trans} for the diagonalization of fermionic quadratic Hamiltonians,
we provide here the analogous prescription for the diagonalization of bosonic quadratic Hamiltonians 
of the general form
\beq \label{eq:Hdiag}
\hat{H}=\sum_{k,m} \left( A_{k m}\hat{b}_k^{\dagger} \hat{b}_{m}
+\frac1{2} B_{k m} ( \hat{b}_k^{\dagger} \hat{b}_{m}^{\dagger}+ \hat{b}_k \hat{b}_{m} )
\right)\,,
\eeq
where $\hat{b}_k$ and $\hat{b}_k^{\dagger}$ are bosonic annihilation and creation operators,
respectively, and $A_{k m}$ and $B_{k m}$ are real-valued and symmetric.
For the spin-wave Hamiltonian of Eq.\ (\ref{eq:Hsw2}), the coefficients are
\begin{subequations}
\begin{align}
A_{k m} &= (D-A \gamma_k) \delta_{k m} + C \delta_{k,m+L/2} \,,\\
B_{k m} &= 2 B \gamma_k \delta_{k, L-m} \,.
\end{align}
\end{subequations}
The diagonalization process starts by defining the following linear transformation
\beq \label{eq:trans}
\hat{\eta}_{\bk} &= \sum_m \left( g_{km} \hat{b}_m +h_{km} \hat{b}_m^{\dagger} \right) \,, 
\eeq
where $g_{km}$ and $h_{km}$ are real-valued and we require $\hat{\eta}_{\bk}$ and $\hat{\eta}_{\bk}^{\dagger}$ be
bosonic operators. This is enforced by the constraint
\beq \label{eq:reqEta}
\delta_{k m}=[\eta_k,\eta_{m}^{\dagger}]=\sum_{l} \left( g_{k l} g_{m l} - h_{k l} h_{m l} \right) \,.
\eeq
The coefficients $g_{k m}$ and $h_{k m}$ are determined in such a way
that the transformed Hamiltonian takes the diagonal form
\beq \label{eq:Heta}
H = \sum_{\bk} \Lambda_{\bk} \hat{\eta}_{\bk}^{\dagger} \hat{\eta}_{\bk}  +E_0 \,,
\eeq
once the  $\hat{\eta}_k$'s are substituted for the $\hat{b}_k$'s,
and \hbox{$E_0= -\sum_{k,m} \Lambda_{m} h_{m k}^2$}.
As the new Hamiltonian is already in diagonal form, the new field operators obey
the eigenvalue equation
\beq
[\hat{\eta}_{\bk},\hat{H}] = \Lambda_{\bk} \hat{\eta}_{\bk} \,.
\eeq
Plugging in the transformations given in Eqs.\ (\ref{eq:trans}), we obtain the relations
\begin{subequations}  \label{eq:Lambdas}
\begin{align}
\Lambda_{\bk} g_{km} &= \sum_l \left( g_{kl} A_{lm} - h_{kl} B_{ml}\right) \,,\\
\Lambda_{\bk} h_{km} &= \sum_l \left( g_{kl} B_{ml} - h_{kl} A_{lm}\right) \,.
\end{align}
\end{subequations}
These relations may be further simplified by defining the new coefficients
\begin{subequations}
\begin{align}
\phi_{km}&= g_{km}+h_{km} \,, \\
\psi_{km}&=g_{km}-h_{km} \,,
\end{align}
\end{subequations}
for which, the constraint (\ref{eq:reqEta}) translates to
\beq
\frac1{2} \sum_l \left( \phi_{kl} \psi_{ml} + \psi_{kl} \phi_{ml} \right) =\delta_{k m}\,.
\eeq
With the above definitions, Eqs.\ (\ref{eq:Lambdas}) may be cast in vector notation:
\begin{subequations}
\begin{align}
\bphi_{\bk} (\bA-\bB) &= \Lambda_{\bk} \bpsi_{\bk} \,, \\
\bpsi_{\bk} (\bA+\bB) &= \Lambda_{\bk} \bphi_{\bk} \,.
\end{align}
\end{subequations}
These can be solved by simply plugging each of these equations into the other, 
resulting in the eigenvalue equations
\begin{subequations}
\begin{align}
\bpsi_{\bk} (\bA+\bB)(\bA-\bB) \, &= \Lambda_{\bk}^2 \bpsi_{\bk} \,, \\
\bphi_{\bk} (\bA-\bB)(\bA+\bB) \, &= \Lambda_{\bk}^2 \bphi_{\bk} \,. 
\end{align}
\end{subequations}
These equations are to be solved by standard techniques.
Once the $\Lambda_{\bk}$'s, $\bphi_{\bk}$'s and $\bpsi_{\bk}$'s are found,
all physical observables can be readily calculated: First, the observable of interest 
should be expressed in terms of normal-ordered $\hat{\eta}_k$'s. 
This may be accomplished by using the inverse of the transformation  given in Eq.\ (\ref{eq:trans}): 
\beq
\hat{b}_k= \frac1{2} \sum_m \Big( (\phi_{km}^{-1} + \psi_{km}^{-1} ) \hat{\eta}_m 
+ 
(\phi_{km}^{-1} - \psi_{km}^{-1}) \hat{\eta}_m^{\dagger} \Big) \,. \nonumber \\
\eeq
As a next step, one should use the fact that as excitations, 
the $\hat{\eta}_{\bk}$'s obey $\hat{\eta}_{\bk}|\GS\rangle_{\mf}=0$. This leads to
\beq
{_{\mf}} \langle \GS| \hat{b}_k^{\dagger} \hat{b}_{m} | \GS \rangle_{\mf}=
\frac1{4} \sum_l \left(  \phi_{kl}^{-1} -  \psi_{kl}^{-1} \right) \left(  \phi_{ml}^{-1} -  \psi_{ml}^{-1}  \right) \,.\nonumber\\
\eeq
As an example, consider the spin-wave corrected density of particles in our model.
It is calculated as
\begin{widetext}
\beq \label{eq:rhoSW}
\rho_{\sw}&=&
\frac1{N} \sum_i {_{\mf}}\langle \GS| \hat{a}_{i}^{\dagger} \hat{a}_{i} | \GS \rangle_{\mf} =
\rho_{\mf} - \frac1{N} \sum_i  {_{\mf}} \langle \GS| \hat{b}_i^{\dagger} \hat{b}_i \cos \theta_i |\GS \rangle_{\mf} 
\\\nonumber 
&=& \rho_{\mf} - \frac1{2 N} \left(\cos \theta_1 + \cos \theta_2  \right) 
\sum_k  {_{\mf}} \langle \GS| \hat{b}_k^{\dagger} \hat{b}_k |\GS \rangle_{\mf} 
-\frac1{2 N} \left(  \cos \theta_1 - \cos \theta_2  \right) 
\sum_k  {_{\mf}} \langle \GS| \hat{b}_k^{\dagger} \hat{b}_{k+L/2} | \GS \rangle_{\mf}  \nonumber\\
&=&
\rho_{\mf} - \frac1{8 N} \left( 
\left(\cos \theta_1 +\cos \theta_2  \right) \sum_{mk} 
\left(  \phi_{km}^{-1} -  \psi_{km}^{-1} \right)^2
+ \left( \cos \theta_1 - \cos \theta_2  \right) \sum_{mk} 
\left(  \phi_{km}^{-1} -  \psi_{km}^{-1} \right) \left(  \phi_{(k+L/2),m}^{-1} -  \psi_{(k+L/2),m}^{-1}  \right)
 \right) \,.\nonumber
\eeq
All other observables may be calculated in the same manner.
\end{widetext}


\begin{thebibliography}{100}
\bibitem{gr1}
M. Greiner, O. Mandel, T. Esslinger, T. E. H\"ansch, and I. Bloch, 
Nature (London) {\bf 415}, 39 (2002).
\bibitem{gr2}
M. Greiner, O. Mandel, T. E. H\"ansch, and I. Bloch, 
Nature (London) {\bf 419}, 51 (2002).
\bibitem{sachdev}
S. Sachdev, {\it Quantum Phase Transitions} (Cambridge University Press, Cambridge, England, 1999).
\bibitem{r4} 
M. P. A. Fisher, P. B. Weichman, G. Grinstein, and D. S. Fisher, 
Phys. Rev. B {\bf 40}, 546 (1989).
\bibitem{r5}
D. Jaksch, C. Bruder, J. I. Cirac, C. W. Gardiner, and P. Zoller, 
Phys. Rev. Lett. {\bf 81}, 3108 (1998).
\bibitem{r6}
W. Zwerger, J. Opt. B: Quantum Semiclassical, 
Opt. {\bf 5}, 9 (2003).
\bibitem{bloch08} 
I. Bloch, J. Dalibard, and W. Zwerger,
Rev. Mod. Phys. {\bf 80}, 885 (2008).
\bibitem{fhb}
E. H. Lieb and F. Y. Wu, Physica A {\bf 321}, 1 (2003).
\bibitem{sce1}
J. K. Freericks and H. Monien, Phys. Rev. B {\bf 53}, 2691 (1996).
\bibitem{sce2}
N. Elstner and H. Monien, Phys. Rev. B {\bf 59}, 12184 (1999).
\bibitem{cg}
A. P. Kampf and G. T. Zimanyi, Phys. Rev. B {\bf 47}, 279 (1993).
\bibitem{mft}
K. Sheshadri, H. R. Krishnamurthy, R. Pandit, and T. V. Ramakrishnan, Europhys. Lett. {\bf22}, 257 (1993).
\bibitem{ftapp}
T. P. Polak and T. K. Kope\'c, Phys. Rev. B {\bf 76}, 094503 (2007). 
\bibitem{per1}
N. Teichmann, D. Hinrichs, M. Holthaus, and A. Eckardt, Phys. Rev. B {\bf 79}, 100503(R) (2009).
\bibitem{Aizenman}
M. Aizenman, E. H. Lieb, R. Seiringer, J. P. Solovej, and J. Yngvason, 
Phys. Rev. A {\bf 70}, 023612 (2004). 
\bibitem{Val}
V. G. Rousseau, D. P. Arovas, M. Rigol, F. H\'ebert, G. G. Batrouni, and R. T. Scalettar,
Phys. Rev. B {\bf 73}, 174516 (2006). 
\bibitem{rigol04} M. Rigol and A. Muramatsu, 
Phys. Rev. A {\bf 70}, 031603(R) (2004); Phys. Rev. A {\bf 72}, 013604 (2005).
\bibitem{rigol06}
M. Rigol, A. Muramatsu, and M. Olshanii,
Phys. Rev. A {\bf 74}, 053616 (2006).
\bibitem{SSE1}
A. W. Sandvik, Phys. Rev. B {\bf 59}, R14157 (1999). 
\bibitem{SSE2}
A. Dorneich and M. Troyer, Phys. Rev. E {\bf 64}, 066701 (2001). 
\bibitem{hcb}
K. Bernardet, G. G. Batrouni, J.-L. Meunier, G. Schmid, M. Troyer and A. Dorneich, 
Phys. Rev. B {\bf 65}, 104519 (2002).
\bibitem{Bruder}
C. Bruder, R. Fazio and G. Sch\"on, Annalen der Physik {\bf 14}, 566 (2005).
\bibitem{peil03}
S. Peil, J. V.~Porto, B.~Laburthe Tolra, J. M.~Obrecht, B. E.~King, M.~Subbotin, 
S. L.~Rolston, and W. D.~Phillips, Phys. Rev. A {\bf 67}, 051603(R) (2003).
\bibitem{strabley06} J. Sebby-Strabley, M. Anderlini, P.~S. Jessen, 
and J.~V. Porto, Phys. Rev. A {\bf 73}, 033605 (2006).
\bibitem{strabley07}
J. Sebby-Strabley, B. L. Brown, M. Anderlini, P. J. Lee, W. D. Phillips, J. V. Porto, and P. R. Johnson,
Phys. Rev. Lett. {\bf 98}, 200405 (2007). 
\bibitem{lee07}
P. J. Lee, M. Anderlini, B. L. Brown, J. Sebby-Strabley, W. D. Phillips, and J. V. Porto,
Phys. Rev. Lett. {\bf 99}, 020402 (2007).
\bibitem{2dcrit}
A. Priyadarshee, S. Chandrasekharan, J.-W. Lee, and H. U. Baranger,
Phys. Rev. Lett. {\bf 97}, 115703 (2006).
\bibitem{mats}
T. Matsubara and H. Matsuda, Prog. Theor. Phys. {\bf 16}, 569 (1956).
\bibitem{rhoS}
M. E. Fisher, M. N. Barber, and D. Jasnow, Phys. Rev. A {\bf 8}, 1111 (1973).
\bibitem{SW1}
K. S. Liu and M. E. Fisher, J. Low Temp. Phys. {\bf 10}, 655 (1973).
\bibitem{SW2}
Yi-Chen Cheng, Phys. Rev. B {\bf 23}, 157 (1981).
\bibitem{SW3}
R. T. Scalettar, G. G. Batrouni, A. P. Kampf, and G. T. Zimanyi, Phys. Rev. B {\bf 51}, 8467 (1995).
\bibitem{SW4}
G. Murthy, D. Arovas, and A. Auerbach, Phys. Rev. B {\bf 55}, 3104 (1997).
\bibitem{trans}
E. Lieb, T. Schultz and D. Mattis, Annals of Physics {\bf 16}, 407 (1961).
\end{thebibliography}
\end{document}